\documentclass[12pt,aps,pra,showpacs,superscriptaddress,showkeys]{revtex4-1}

\usepackage{color,stmaryrd,amsmath,amsthm,amsfonts,amssymb,times,bbm,graphicx,tikz,verbatim}

\usetikzlibrary{arrows,fadings,decorations}

\usepackage{amsmath,amsfonts,amssymb,bbm}

\newcommand{\acco}[1]{\left\{#1\right\}}
\newcommand{\bra}[1]{\langle #1 |}

\newcommand{\pa}[1]{\left(#1\right)}
\newcommand{\ket}[1]{| #1 \rangle}

\newcommand{\incl}{\subseteq}

\newcommand{\joliA}{\mathcal{A}}

\newcommand{\joliC}{\mathcal{C}}

\newcommand{\joliH}{\mathcal{H}}

\newcommand{\joliN}{\mathcal{N}}

\newcommand{\Z}{\mathbb{Z}}

\newcommand{\id}{\operatorname{id}}

\def\qneig#1{\joliN(Q_{#1})}
\def\joliNT{\joliN^T}
\def\BN{\mathcal{BN}}
\newcommand{\tof}{f}

\newcommand\tofcel[4]{\fill[color=gray] (#1,#2) ellipse (1.225cm and 0.8cm);
\fill[white] (#1+0.625,#2) circle (0.5cm);
\fill[white] (#1-0.625,#2) circle (0.5cm);
\draw (#1-0.625,#2) node[circle] {#3};
\draw (#1+0.625,#2) node[circle] {#4};
}

\newcommand\linecircuit[2]{\draw (#1,#2) -- (#1+1.5,#2+3);}

\newcommand\transvershaut[2]{

\draw (#1+0.9,#2+1.8) -- (#1+0.9+2.5,#2+1.8);
\draw (#1+0.9+1.25,#2+1.8) circle (0.15) ;
\draw[fill] (#1+0.9+2.5,#2+1.8) circle (0.1) ;
\draw[fill] (#1+0.9,#2+1.8) circle (0.1) ;
}
\newcommand\transversbas[2]{\draw (#1+0.6,#2+1.2) -- (#1+0.6+2.5,#2+1.2);
\draw (#1+0.6+1.25,#2+1.2) circle (0.15) ;
\draw[fill] (#1+0.6+2.5,#2+1.2) circle (0.1) ;
\draw[fill] (#1+0.6,#2+1.2) circle (0.1) ;
}

\newtheorem{prop}{Proposition}

\newtheorem{cor}{Corollary}

\def\xorca{{\textsc XOR-CA}}
\def\toffca{{\textsc Toffoli-CA}}
\def\qi{\hbox{\bf\Large\_\,}}
\def\qq{\qi\!\qi}

\begin{document}

\pgfdeclaredecoration{complete sines}{initial}
{
    \state{initial}[
        width=+0pt,
        next state=sine,
        persistent precomputation={\pgfmathsetmacro\matchinglength{
            \pgfdecoratedinputsegmentlength / int(\pgfdecoratedinputsegmentlength/\pgfdecorationsegmentlength)}
            \setlength{\pgfdecorationsegmentlength}{\matchinglength pt}
        }] {}
    \state{sine}[width=\pgfdecorationsegmentlength]{
        \pgfpathsine{\pgfpoint{0.25\pgfdecorationsegmentlength}{0.5\pgfdecorationsegmentamplitude}}
        \pgfpathcosine{\pgfpoint{0.25\pgfdecorationsegmentlength}{-0.5\pgfdecorationsegmentamplitude}}
        \pgfpathsine{\pgfpoint{0.25\pgfdecorationsegmentlength}{-0.5\pgfdecorationsegmentamplitude}}
        \pgfpathcosine{\pgfpoint{0.25\pgfdecorationsegmentlength}{0.5\pgfdecorationsegmentamplitude}}
}
    \state{final}{}
}

\title{Bounds on the Speedup in Quantum signalling}

\author{Pablo Arrighi}
\affiliation{
Aix-Marseille Univ., LIF, F-13288 Marseille Cedex 9, France}
\email{pablo.arrighi@univ-amu.fr}, 

\author{Vincent Nesme}
\affiliation{LIG, Universit\'e Joseph Fourier, Grenoble, France}
\email{vincent.nesme@imag.fr}

\author{Reinhard F. Werner}
\affiliation{Institut f\"ur Theoretische Physik, Leibniz Universit\"at, Hannover, Germany}


\begin{abstract}
Given a discrete reversible dynamics, we can define a quantum dynamics, which acts on basis states like the classical one, but also allows for superpositions of them. It is a curious fact that in the quantum version, local changes in the initial state, after a single dynamical step, can sometimes can be detected much farther away than classically.  Here we show that this effect is no use for generating faster signals. In a run of many steps the quantum propagation neighborhood can only increase by a constant fringe, so there is no asymptotic increase in speed.
\end{abstract}

\maketitle

\section{Introduction}

It is well known that a unitary quantum walk on a lattice propagates much faster than its classical counterpart: where the classically walking particle is expected to travel a distance of the order of $\sqrt t$ in $t$ steps, the quantum particle travels ballistically, covering a distance of the order of $t$. The comparison is not entirely fair, because as soon as randomness in the form of decoherence is introduced the quantum walk also slows down to diffusive scaling. In this paper, as in \cite{thooft}, we study the complementary fair case, namely propagation in a non-random, fully reversible classical dynamical system and its quantum counterpart.

The classical dynamical system here will be a Cellular Automaton. Cellular automata (CA), first introduced by von Neumann \cite{Neumann}, consist of an array of identical cells, each of which may take one of a finite number of possible states. The entire array evolves in discrete time steps by applying the same local transition function everywhere, synchronously. CA are used in Computer Science to model space-sensitive problems such as self-reproduction or synchronization, but they also arise quite naturally in applied mathematics and physics, as discrete models and numerical schemes for PDEs. In the context of this paper the cells need not be arranged in a grid, and the local transition function need not be the same everywhere: by CA we really mean just discrete time discrete space dynamics. 

The quantum counterpart of this dynamical system will similarly be a Quantum Cellular Automaton (QCA). Indeed, because CA are a physics-like model of computation \cite{MargolusPhysics}, Feynman \cite{FeynmanQCA}, and later Margolus \cite{MargolusQCA}, suggested early in the development of Quantum Computing that quantizing this model was important. For two main reasons. First, they are a natural framework in which to cast the quantum simulation of a quantum system \cite{Bialynicki-Birula,MeyerQLGI}. Secondly, because they seem advantageous as an implementation architecture for a quantum computer \cite{LloydQCA,VollbrechtCirac}: in QCA computation occurs without extraneous (unnecessary) control, hence eliminating a source of noise. There are yet other reasons to study QCA: as a model of distributed quantum computation, as a mathematical framework in order to grasp the interplay between entanglement and causality \cite{ArrighiUCAUSAL,ArrighiJCSS}\cite{GUWZ}, or even as yielding toy models of quantum spacetime \cite{LloydQG}. QCA are the natural multi-particle extensions of Quantum Walks. But again, in the context of this paper the cells do not need to be arranged in a grid, and the evolution does not need to act the same everywhere: by QCA we really mean just a discrete time discrete space quantum dynamics. Actually, since we specifically look at QCA arising from quantizing CA, their state space will be akin to a Hilbert space whose basis states are labelled by the configurations of the classical CA, and their unitary dynamics will be completely determined by the by the linear extension of the classical CA.

In both the quantum and classical settings there is a clear notion of localization, i.e., observables associated to each cell or a group of cells, and therefore we can directly compare propagation properties. We will see that a large quantum speedup is possible, in the sense that the quantum propagation neighbourhood can be much larger than the classical one. On the other hand we show that this gain cannot be realized in every step: in the long run of $t$ steps the neighborhood gain does not increase with $t$. Therefore, the asymptotic speed of propagation is the same in QCA and CA.

The first upper bound on the quantum propagation neighborhood was given in \cite{Schumacher} (compare to (\ref{onestep}) below) in the context of proving that the above procedure for ``quantizing'' a CA is well-defined. However, no examples showing the quality of the bound were given. The theory was further developed in \cite{ANW1,block}. In particular, an interpretation of the quantum neighborhood in terms of the classical CA, called the ``block neighborhood'' was given in \cite{block}. Our bound for the asymptotic case crucially depends on the bound on block neighborhoods derived there.

Our paper is organized as follows. We begin with an example, the \xorca, which shows, paradoxically, an infinite speedup. Although the classical dynamics is local, the quantum system can carry signals arbitrarily far in a single step. In the definition of \cite{Schumacher} the quantum system is thus not a QCA. The analysis of this extreme case helps us to explain the origin of the speedups discussed in this paper, and also highlights the importance of the classical inverse neighborhood (which is infinite in this example). We then make our mathematical setting precise, in particular excluding such pathologies and giving a more formal definition of the neighborhoods. We show in our second basic example, the \toffca, a quantum neighborhood twice as large as the classical one. We then state the best known bounds on the single-step quantum neighborhood. The effect of asymptotic loss of speedup is also first explained in the example of the \toffca, accompanied by a general theorem to this effect.

\section{The \xorca: infinite speedup?} 
A Cellular Automaton (CA) is a function from configurations to configurations. Configurations are themselves functions, which associate, to each point of the lattice, a state in a finite set. In our first example we take the lattice to be ${\cal Z}$, and the finite set to be $\Sigma =\{\qi,0,1\}$. The set of infinite configurations ${\cal C}$ is then the set of functions from ${\cal Z}$ to $\Sigma$. 
The set of finite configurations ${\cal C}_f$ is more restricted: it contains only those configurations $c=\cdots c_{-1}c_0c_1c_2\cdots$ such that there are only a finite number of $c_i\neq\qi$. The symbol $\qi$ is called the quiescent state. This restriction to finite configurations is a standard assumption for the CA and Turing machines computational models, to exclude infinite parallel computation or uncomputable input configurations.
The dynamical transformation $f$ is given in terms of a local transition rule $\delta$ applied to the contents of each cell and its right neighbor. It is  based on the $\textsc{xor}$ gate, i.e., addition mod 2, which we denote by $\oplus$: for $x,y\in\acco{0,1}$, $\delta(\qi x)=\qi$, $\delta(x\qi)=x,$ and $\delta(xy)=x\oplus y$. The induced, global, classical dynamics $f$ thus takes the configuration $c=\cdots c_{i-1} c_i c_{i+1} \cdots$ to the configuration $f(c)=\cdots \delta(c_{i-1} c_i)\delta(c_{i} c_{i+1})\cdots$. The infinite configurations $\ldots 000 \ldots$ and $\ldots 111 \ldots$ have the same image, namely $\ldots 000\ldots$, hence $f$ is not bijective over ${\cal C}$ the set of infinite configurations. However, one can easily work out that it is bijective over the set of finite configurations ${\cal C}_f$. Hence its quantization $Q_f$, or ``linear extension'' defined on basis states as $Q_f\ket{c}=\ket{f(c)}$, is a perfectly valid unitary operator from ${\cal H}_{\mathcal{C}_f}$ to ${\cal H}_{\mathcal{C}_f}$, where ${\cal H}_{\mathcal{C}_f}$ is the Hilbert space having $\mathcal{C}_f$ as its orthonormal basis. So $f$ is a well-defined cellular automaton, and everything seems to suggest that $Q_f$ is a perfectly valid quantization of $f$, i.e. a quantum cellular automaton (QCA).  But no: surprisingly $Q_f$ violates the causality condition for QCAs in \cite{Schumacher}. That is, it allows to transmit information arbitrarily far in a single step \cite{ANW1}.

Indeed, for $x=0$ or $1$, consider the configuration $d^x=\cdots \qq00\cdots 0x\qq\cdots$, where the dots in the middle stand for some long segment of the lattice, say of size $L$. It has antecedent $c^x=f^{-1}(d^x)=\cdots\qq xx\cdots xx\qq\cdots$. Now consider the two superposition states $\ket{c^\pm}=(\ket{c^0}\pm\ket{c^1})/\sqrt2$  and their images $\ket{d^\pm}=Q_f\ket{c^\pm}$:
\begin{eqnarray}
  \ket{c^\pm}&=&{\textstyle\frac{1}{\sqrt{2}}}\ket{\cdots\qq}\otimes \bigl(\ket{00\cdots 00}\pm\ket{11\cdots 11}\bigr)\otimes \ket{\qq\cdots}
            \nonumber\\
   \ket{d^\pm}&=&\ket{\cdots\qq00\cdots 0}\otimes \ket{\pm}\otimes \ket{\qq\cdots}\nonumber
\end{eqnarray}
where we used the usual notation $\ket{\pm}=\frac{1}{\sqrt{2}}(\ket{0}\pm\ket{1})$.

Let us now describe how one can transmit information between arbitrarily distant parties in just one step of this dynamics.  The line is prepared in the state $c_+$ with the first non quiescent cell in  Alice's lab in Paris and the last non quiescent cell with Bob in New York. Then Alice either leaves the state unchanged or performs a \emph{local change} by applying a phase gate $Z$ to her cell, changing $c^{+}$ into $c^{-}$. Then one \xorca\ is performed leading to either $\ket{d^+}$ or $\ket{d^-}$, and hence  a perfectly measurable change from $\ket{+}$ to $\ket{-}$ for Bob, despite him being arbitrarily far remote.

This infinite speedup is intuitively unphysical. To make this intuition precise let us try to implement the \xorca\  with local gates satisfying the natural constraints that (1) each gate operates on a well defined subset of the system qubits and possibly some locally available ancillas, (2) in each clock cycle these subsets of simultaneously operating gates do not overlap, to avoid double use of quantum information and hence illegal cloning and (3) there are finitely many clock cycles. Any automaton built in this way is {\it structurally reversible}: we can invert it, by inverting the steps in each clock cycle, and the inverse is again a cellular automaton. This is precisely what the problem with this \xorca: although $f$ has an inverse on finite configurations, this inverse is not itself a CA, i.e. there is no upper bound on the number of cells one has to look at in order to compute the antecedent. In fact, the same situation always arises whenever a CA $f$ is one-to-one on finite configurations, but fails to be one-to-one on infinite configurations. This is because of a deep theorem in CA theory \cite{Hedlund}, which relies on the compactness of ${\cal C}$ equipped with a certain metric, and the characterization of CA as continuous functions with respect to that metric, in order to prove that if $f$ is a CA over ${\cal C}$ and has an inverse, then $f^{-1}$ is itself a CA over ${\cal C}$.\\
We will assume in the rest of the paper that both $f$ and $f^{-1}$ are CAs, i.e., have finite neighborhoods $\joliN(f)$ and $\joliN(f^{-1})$, and we will provide lower and upper bounds for the neighborhoods of the corresponding $Q_f$. In the lower bound both $\joliN(f)$ and $\joliN(f^{-1})$ appear. Porting this lower bound argument back to the $\joliN(f^{-1})$ infinite case, implies that the above \xorca\ trick can always be applied. In other words, any CA which is bijective over finite configurations but not over infinite configurations, has a infinite $\joliN(f^{-1})$, and hence an infinite $\joliN(Q_f)$.

\section{Setting and one-step bound} 

\subsection{Classical state space and evolutions} 

For the general points we want to make the structure of the lattice $X$ which labels the cells is largely irrelevant. Although our examples are drawn from one-dimensional lattices, any dimension is fine, and translation invariance is also not needed. 
We only require that the system be organized in a set $X$ of cells, and that the classical content of cell $x\in X$ be taken from a finite alphabet $\Sigma_x$, which may depend on $x$.  For any subset $A$ of $X$, we denote by $\Sigma_A=\prod_{x\in A}\Sigma_x$ the set of configurations on $A$.  When $A$ is not mentioned, it is understood to be the whole set $X$, so that a configuration $c$ is an element of $\joliC=\Sigma_X$; a classical evolution is a function $f:\joliC\to\joliC$.

\subsection{Neighborhoods}

Neighborhoods $\joliN(.)$ are defined in an operational way which applies to the classical and quantum side alike, namely as the set of pairs $(x,y)$ such that a state change at $x$ after one step of the dynamics can make a detectable difference at $y$. To any evolution $f$ over classical configurations, we can thus associate a neighborhood $\joliN(f)$, so that $(x,y)\in \joliN(f)$ if and only if there exists two configurations $c$, $d$ which differ only at cell $x$ and such that $f(c)_y\neq f(d)_y $.  Then an evolution $f$ is called a (non translation-invariant) {\em cellular automaton} (CA for short) if, for all $y$, the set $\joliN(f)(y)=\{x|(x,y)\in\joliN(f)\}$  is finite, so that the update of cell $y$ can be computed from finitely many $c_x$. 

\begin{figure}[htbp]
\begin{tikzpicture}
\tikzstyle{gloubi}=[circle, draw, fill=black!50,
                        inner sep=0pt, minimum width=4pt]

\draw (0,0) ellipse  (1.6cm and 0.5cm);
\draw (0,2) ellipse  (1.6cm and 0.5cm);

\draw (0,0) node[gloubi] {} -- (0,2) node[gloubi] {};

\draw (0.3,0) node {$x$} ;

\draw (0.3,2) node {$y$} ;
\draw[->=latex,thick] (-1.5,0.2) to [bend left=30] (-1.5,1.8) ;

\draw (-2,1) node {$f$} ;

\end{tikzpicture}
\caption{\label{fig:voisinage} $(x,y)\in\joliN(f)$ means that, in at least one case, knowing $c_x$ is essential to determining $f(c)_y$.  In other words, it means that there exists an antecedent configuration $c$ such that changing the cell $x$ will change the cell $y$ of the image configuration $f(c)$.}
\end{figure}
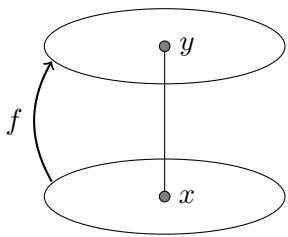

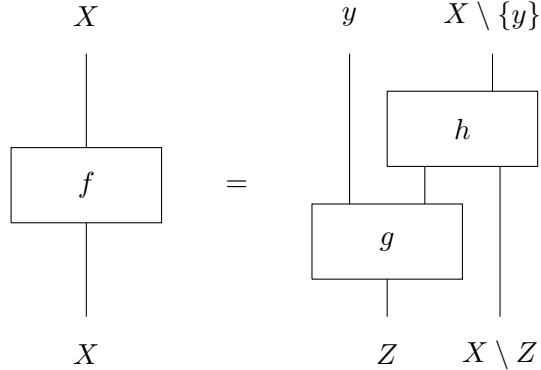
\begin{figure}[htbp]
\begin{tikzpicture}[scale=2]
\tikzstyle{gloubi}=[circle, draw, fill=black!50,
                        inner sep=0pt, minimum width=4pt]

\draw(-1.5,0.25) -- (-1.5,0.875) ;
\draw (-2,0.875) -- (-2,1.375) -- (-1,1.375) -- (-1,0.875) -- (-2,0.875) ;
\draw(-1.5,1.375) -- (-1.5,2) ;
\draw (-1.5,1.125) node {$f$} ;

\draw (-1.5,0) node {$X$} ;
\draw (-1.5,2.25) node {$X$} ;

\draw (-0.5,1.125) node {$=$} ;

\draw (0.5,0.25) -- (0.5,0.5) ;
\draw (0,0.5) -- (1,0.5) -- (1,1) -- (0,1) -- (0,0.5) ;
\draw (0.5,0.75) node {$g$} ;
\draw (0.25,1) -- (0.25,2) ;
\draw (0.75,1) -- (0.75,1.25) ;
\draw (0.5,1.25) -- (1.5,1.25) -- (1.5,1.75) -- (0.5,1.75) -- (0.5,1.25) ;
\draw (1,1.5) node {$h$} ;
\draw (1.25,0.25) -- (1.25,1.25) ;
\draw (1.2,1.75) -- (1.2,2) ;
\draw (0.25,2.25) node {$y$} ;
\draw (1.2,2.25) node {$X\setminus\acco{y}$} ;

\draw (0.5,0) node {$Z$} ;
\draw (1.25,0) node {$X\setminus Z$} ;

\end{tikzpicture}
\caption{\label{fig:semilocal} $\BN(f)(y)$ is the smallest $Z\incl X$ such that $f$ can be semilocalized, i.e. decomposed in such a way, with $g$ and $h$ bijective.  Note that $X$ and $Z$ denote neither antecedents nor images of $f$; they are sets of cells.  For instance, the left hand side reads ``the image of a configuration on $X$ by $f$ is a configuration on $X$'' and not ``$f(X)=X$''.}
\end{figure}

In this paper, we assume that $f$ is one-to-one, and that both $f$ and $f^{-1}$ are CA.  Under these assumptions, another neighborhood scheme introduced in \cite{ANW1}, namely the Block Neighborhood $\BN(.)$, is also finite.  This notion arises naturally when we demand that the local mechanism that implements $f$ be itself one-to-one, and wonder about its minimal size. To any evolution $f$ of the classical configurations, we can thus associate a block neighborhood $\BN(f)$, so that ``$(x,y)\in \BN(f)$'' translates to ``$x$ is in the range of minimal local reversible gate which computes $y$'', for the dynamics $f$. Formally, the definition of $\BN(f)(x)$ is given in Fig. \ref{fig:semilocal}.

\subsection{Quantum state space and evolutions} 

To each $x\in X$ we associate a Hilbert space $\joliH_x$, endowed with an orthonormal basis $\acco{\ket{a} | a\in \Sigma_x}$.  The local observable algebra associated to a cell $x$ is $\joliA_x=L(\joliH_x)$.  To each finite subset $I$ of $X$ we associate $\joliA_I=\bigotimes_{x\in I} \joliA_x$; if $I\incl J$, there is a natural embedding of $\joliA_I$ into $\joliA_J$.  The limit of this system of inclusions is called the local algebra, and denoted $\joliA$; for all practical purposes an element $A\in\joliA$ can be thought of as a local operation $A=A_I\otimes Id_{X\setminus I}$ with $I$ a finite subset of $X$ \citep{Bratteli}.  A quantum evolution $Q$ is just an automorphism $Q:\joliA\to\joliA$, i.e. a linear operator such that $Q(AB)=Q(A)Q(B)$ \cite{Schumacher}, but a more concrete view is to say that it maps any $A\in\joliA$ into $Q(A)= U^\dagger AU\in\joliA$ for some unitary operator $U$. Informally, this $U$ can be thought of as acting on ``$\bigotimes_{x\in X}\joliH_x$'', i.e. it evolves the superpositions of configurations, in the Shr\"odinger picture. Again to any quantum evolution $Q$, we can associate a neighborhood $\joliN(Q)$, so that ``$(x,y)\in \joliN(Q)$'' translates to ``$x$ can influence $y$'', for the quantum evolution $Q$. Formally, by definition, $(x,y)\not\in\joliN(Q)$ iff $Q(\joliA_y)\incl \joliA_{X\setminus\acco{x}}$.


\subsection{Quantization}

In this paper, we are interested specifically in the quantum evolutions $Q_f$ obtained by quantizing an evolution $f$.  Intuitively, such a $Q_f$ arises as follows. First, consider $U_f$ the linear extension of $f$, which maps $\sum_i \alpha_i \ket{c^i}$ into $\sum_i \alpha_i \ket{f(c^i)}$. Then, let $Q_f(A)$ be ${U_f}^\dagger A U_f$. The problem with this approach is that $U_f$ is not itself a member of $\joliA$, and so it is not clear whether $Q_f$ makes sense as an operator over $\joliA$. In fact this fails to be the case for the \xorca.

\begin{figure}[htbp]
\begin{tikzpicture}[scale=2]
\tikzstyle{gloubi}=[circle, draw, fill=black!50,
                        inner sep=0pt, minimum width=4pt]

\draw (0.5,0.25) -- (0.5,0.5) ;
\draw (0,0.5) -- (1,0.5) -- (1,1) -- (0,1) -- (0,0.5) ;
\draw (0.5,0.75) node {$g^{-1}$} ;
\draw (0.25,1) -- (0.25,2) ;
\draw (0.75,1) -- (0.75,1.25) ;
\draw (0.5,1.25) -- (1.5,1.25) -- (1.5,1.75) -- (0.5,1.75) -- (0.5,1.25) ;
\draw (1,1.5) node {$h^{-1}$} ;
\draw (1.25,0.25) -- (1.25,1.25) ;
\draw (1,1.75) -- (1,2.75) ;

\draw (0,2) -- (0.5,2) -- (0.5,2.5) -- (0,2.5) -- (0,2) ;
\draw (0.25,2.25) node {$A$} ;
\draw (0.5,2.75) -- (1.5,2.75) -- (1.5,3.25) -- (0.5,3.25) -- (0.5,2.75) ;
\draw (1,3) node {$h$} ;

\draw (0.25,2.5) -- (0.25,3.5) ;
\draw (0,3.5) -- (1,3.5) -- (1,4) -- (0,4) -- (0,3.5) ;
\draw (0.5,3.75) node {$g$} ;

\draw (0.75,3.25) -- (0.75,3.5) ;
\draw (0.5,4) -- (0.5,4.25) ;
\draw (1.25,3.25) -- (1.25,4.25) ;

\draw (2,2.25) node {$=$} ;

\draw (3,0.25) -- (3,0.5) ;
\draw (2.5,0.5) -- (3.5,0.5) -- (3.5,1) -- (2.5,1) -- (2.5,0.5) ;
\draw (3,0.75) node {$g^{-1}$} ;
\draw (2.75,1) -- (2.75,2) ;
\draw (3.25,1) -- (3.25,3.5) ;

\draw (3.75,0.25) -- (3.75,4.25) ;

\draw (2.5,2) -- (3,2) -- (3,2.5) -- (2.5,2.5) -- (2.5,2) ;
\draw (2.75,2.25) node {$A$} ;

\draw (2.75,2.5) -- (2.75,3.5) ;
\draw (2.5,3.5) -- (3.5,3.5) -- (3.5,4) -- (2.5,4) -- (2.5,3.5) ;
\draw (3,3.75) node {$g$} ;

\draw (3,4) -- (3,4.25) ;
\draw (3.75,3.25) -- (3.75,4.25) ;

\draw (0.5,4.5) node {$\BN(f)(y)$} ;
\draw (0.5,0) node {$\BN(f)(y)$} ;
\draw (3,4.5) node {$\BN(f)(y)$} ;
\draw (3,0) node {$\BN(f)(y)$} ;

\draw (4.25,2.25) node {$=$} ;

\draw (5.5,0.25) -- (5.5,2) ;
\draw (4.75,2) -- (6.25,2) -- (6.25,2.5) -- (4.75,2.5) -- (4.75,2) ;
\draw (5.5,2.25) node {$Q_f(A)$} ;
\draw (5.5,2.5) -- (5.5,4.25) ;

\draw (6.5,0.25) -- (6.5,4.25) ;
\draw (5.5,4.5) node {$\BN(f)(y)$} ;
\draw (5.5,0) node {$\BN(f)(y)$} ;

\end{tikzpicture}
\caption{\label{fig:qneig} Let $A$ act on $y$. Intuitively $Q_f(A)$ is ${U_f}^\dagger A U_f$, but this is not always local. However if $\BN(f)(y)$ is finite, $U_f$ decomposes into $U_g$ and $U_h$, and ${U_f}^\dagger A U_f$ yields the left picture. The $U_h$ cancel out (middle), and the remainder ${U_g}^\dagger A U_g$ makes for a good, local definition of $Q_f(A)$.}
\end{figure}
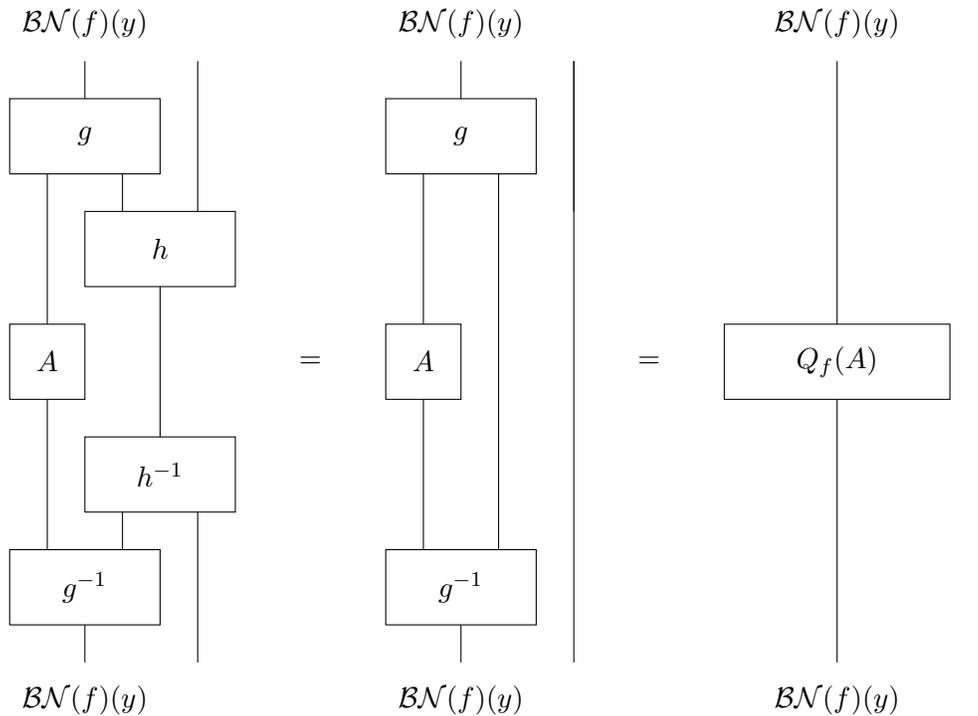

In order reach a rigorous definition, we rely on the assumptions that not only $f$ is bijective, but also that $\joliN(f)(x)$ and $\joliN(f^{-1})(x)$ are finite for every $x\in X$, so that $f$ has finite block neighborhood $\BN(f)$ \citep{ANW1}. Then $Q_f$ will be well-defined, as it will map elements of $\joliA_{y}$ into elements of $\joliA_{\BN(f)(y)}$ and be an automorphism. More precisely, given some $A$ in $\joliA_{y}$, consider the decomposition of $f$ into bijections $g$ and $h$ with $g$ over $\BN(f)(y)$ as in Fig. \ref{fig:semilocal}, and let $U_g$ be the linear extension of $g$. We define $Q_f(A)$ to be ${U_g}^\dagger A U_g$, which is an element of $\joliA_{\BN(f)(y)}$ since so is $U_g$. Next we extend $Q_f$ to the whole of $\joliA$ by demanding that it be an automorphism. Notice that this definition is consistent with the intuition that $Q_f(A)$ be ${U_f}^\dagger A U_f$, since for $A$ in $\joliA_{y}$, we have ${U_f}^\dagger A U_f={U_g}^\dagger A U_g$ as is clearly shown by Fig. \ref{fig:qneig}. This definition of quantization coincides with that of \cite{Schumacher}.

\subsection{Quantum versus block neighborhoods}

Figure~\ref{fig:qneig} provides a definition of $Q_f$, but it also gives a justification that $\qneig{f}$, the \emph{quantum neighborhood} of $f$, is included in $\BN(f)$.  Moreover, as is proven in the appendix, we have the converse inclusion $\qneig{f}\supseteq \BN(f)$.  Quantum and block neighborhoods are thus, as it was hinted in \cite{block} but never actually proven, the same thing.

\begin{prop} \label{prop:main} Let $f:\joliC\to\joliC$ be bijective and such that for every $x\in X$, both $\joliN(f)(x)$ and $\joliN(f^{-1})(x)$ are finite. Then its quantization $Q_f:\joliA\to\joliA$ fulfills $\qneig{f}=\BN(f)$.
\end{prop}

Compare this with \cite{semicausal}, which states that semicausality (i.e. ``the system $y$ can only be influenced by $\joliN(Q)(Y)$'') and semilocalizability (i.e. ``the system $y$ can be computed by a circuit of automorphisms of the form of Fig. \ref{fig:semilocal}'') are equivalent in the quantum regime.  Our proposition is very closely related to that statement; more precisely, it can be seen as its classical counterpart, although neither result is directly derived from the other.  In the remainder of this paper, it will allow us to generalize the results of \cite{block} to the quantum setting, as corollaries.

\subsection{Bounds on the Quantum Neighborhood}

To state the bounds, we introduce a composition of neighborhood sets, by which $(x,z)\in\joliN_1\joliN_2$ means that for some $y$ we have $(x,y)\in\joliN_1$ and $(y,z)\in\joliN_2$. If the $\joliN_i$ are the graphs of functions, this is just the composition of functions. Moreover, this operation matches the composition of automata, so that $\joliN(fg)\incl\joliN(g)\joliN(f)$. By $\joliNT$ we denote the transpose, i.e., the set of pairs $(y,x)$ with $(x,y)\in\joliN$.  Notice that the transposition, as it does with matrices, reverses the order of composition : $(\joliN_1\joliN_2)^{T}=\joliNT_2\joliNT_1$.

Lemma~4 of \cite{Schumacher} can then be expressed as such: 
\begin{equation}\label{onestep}
   \joliN(f)\incl\qneig f \incl\joliN(f)\joliNT(f)\joliNT(f^{-1})\ .
\end{equation}
A crucial property of quantum neighborhoods is that $\joliN(Q^{-1})=\joliNT(Q)$ \cite{ANW3}. This is somewhat surprising, since in the classical case there is not even a bound on the size of $\joliN(f^{-1})$ in terms of $\joliN(f)$. Indeed this is the feature that makes examples like the \xorca\ possible, and makes the problem whether or not an automaton is reversible undecidable in $\geq2$ dimensions. In contrast, computing the quantum inverse is literally as easy as transposing and conjugating a unitary matrix.

If we apply (\ref{onestep}) to $f^{-1}$ we get 

\begin{equation}
   \joliN(f^{-1})\incl\qneig {f^{-1}} \incl\joliN(f^{-1})\joliNT(f^{-1})\joliNT(f)\ .
\end{equation}

The quantization $Q_{f^{-1}}$ of the inverse of $f$ is equal to $Q_{f}^{-1}$, the inverse operation of the quantization of $f$.  We can therefore write

\begin{equation}
   \joliN(f^{-1})\incl\joliN (Q_f^{-1}) \incl\joliN(f^{-1})\joliNT(f^{-1})\joliNT(f)\ .
\end{equation}

We now take the transpose of these terms; keeping in mind that this operation reverses the order of composition, we get

\begin{equation}
   \joliNT(f^{-1})\incl\joliNT (Q_f^{-1}) \incl \joliN(f)\joliN(f^{-1})\joliNT(f^{-1})\ .
\end{equation}

Lastly, since $\joliN(Q^{-1})=\joliNT(Q)$, we have 

\begin{equation}\label{onestepDual}
   \joliNT(f^{-1})\incl\qneig f \incl \joliN(f)\joliN(f^{-1})\joliNT(f^{-1})\ .
\end{equation}

\begin{cor}\label{prop:onestepTotal}
$\joliN(f)\cup\joliNT(f^{-1})\incl \qneig f \incl \pa{\joliN(f)\joliNT(f)\joliNT(f^{-1})}\cap \pa{\joliN(f)\joliN(f^{-1})\joliNT(f^{-1})}$.
\end{cor}

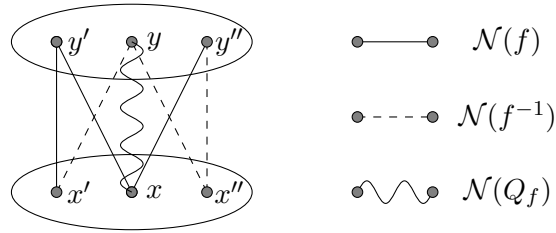
\begin{figure}[htbp]
\begin{tikzpicture}
\tikzstyle{gloubi}=[circle, draw, fill=black!50,
                        inner sep=0pt, minimum width=4pt]

\draw (0,0) ellipse  (1.6cm and 0.5cm);
\draw (0,2) ellipse  (1.6cm and 0.5cm);

\draw[dashed] (-1,0) -- (0,2) ;
\draw[dashed] (1,0)  -- (1,2) ;
\draw[dashed] (1,0) -- (0,2);

\draw (0,0) node[gloubi] {} -- (1,2) node[gloubi] {};
\draw (0,0) node[gloubi] {} -- (-1,2) node[gloubi] {};
\draw (-1,0) node[gloubi] {} -- (-1,2) node[gloubi] {};

\draw (0,2) node[gloubi] {} ;
\draw (1,0) node[gloubi] {} ;

\draw[decoration={
            complete sines,
            segment length=0.5cm,
            amplitude=0.3cm
        },
        decorate] 
 (0,0) -- (0,2);

\draw (3,2) node[gloubi] {} -- (4,2) node[gloubi] {} ;
\draw (5,2) node {$\joliN(f)$} ;

\draw[dashed] (3,1) -- (4,1) ;
\draw (3,1) node[gloubi] {};
\draw (4,1) node[gloubi] {};
\draw (5,1) node {$\joliN(f^{-1})$} ;

\draw%
[decoration={
            complete sines,
            segment length=0.5cm,
            amplitude=0.3cm
        },
        decorate]
(3,0) -- (4,0) ;

\draw (3,0) node[gloubi] {};
\draw (4,0) node[gloubi] {};
\draw (5,0) node {$\qneig f$} ;

\draw (0.3,0) node {$x$} ;
\draw (-0.7,0) node {$x'$} ;
\draw (1.3,0) node {$x''$} ;
\draw (0.3,2) node {$y$} ;
\draw (-0.7,2) node {$y'$} ;
\draw (1.3,2) node {$y''$} ;

\end{tikzpicture}
\caption{\label{fig:onestepthm} Illustration of the combined upper bounds (\ref{onestep}) and (\ref{onestepDual}).  In order to be able to send a signal from $x$ to $y$ in the quantum regime, it is necessary that there exist $x'$, $x''$, $y'$ and $y''$ forming such a pattern (points on each side need not be distinct).  
}
\end{figure}

These are the best general bounds on $\qneig f$ known to us. The dependencies in the classical neighborhoods which must be satisfied for a dependence $(x,y)\in\qneig f$ to occur are shown in Fig.~\ref{fig:onestepthm}.

\section{\toffca:  an example with large quantum step size}
Examples using the ideas of the \xorca, i.e., examples with $\joliN(f^{-1})$ much larger than $\joliN(f)$ show that $\qneig f$ can be much larger than $\joliN(f)$. But can the upper bounds (\ref{onestep}) and (\ref{onestepDual}) also be exhausted? That is, even if we accept for a fact that the inverse neighborhood of $f$ enters $\qneig f$, can we get a speedup? An example, the \toffca\cite{ANW1},  is illustrated in Figure~\ref{fig:Toffoli}.  It is based on the {\sc Toffoli} gate, a double conditioned {\textsc CNOT}. In the diagrams the conditioning is represented as a horizontal line, and the qubits by slanted lines. Since neighboring {\sc Toffoli} gates commute, their ordering is irrelevant. Each single cell now contains two qubits. The
alphabet is $\{00,01,01,11\}$, with $00$ taken as the quiescent symbol. The classical transition function acting for updating a cell with content $cd$, with left neighbor containing $ab$ is $\delta(abcd)= b(c\oplus b\cdot d)$, where the product $b\cdot d$ is just the {\sc And} of bits.

\begin{figure}[htbp]
\begin{tikzpicture}[scale=1.2]
\clip (-1.5,-1) rectangle (6.5,4);

\linecircuit{-0.625}{0}
\linecircuit{0.625}{0}
\linecircuit{2.5-0.625}{0}
\linecircuit{2.5+0.625}{0}
\linecircuit{5-0.625}{0}
\transvershaut{1.25-0.625}{0}
\transversbas{1.25+2.5-0.625}{0}

\linecircuit{5+0.625}{0}
\linecircuit{-2.5+0.625}{0}
\transvershaut{1.25+5-0.625}{0}
\transversbas{1.25-2.5-0.625}{0}
\transvershaut{1.25-5-0.625}{0}

\tofcel{0}{0}{$a$}{$b$}
\tofcel{2.5}{0}{$c$}{$d$}
\tofcel{5}{0}{}{}

\tofcel{0}{3}{}{}
\tofcel{2.5}{3}{$b$}{\begingroup \renewcommand*{\arraystretch}{.5} $\begin{array}{c} c\\ \oplus \\ b\cdot d \end{array}$\endgroup}
\tofcel{5}{3}{}{}
\end{tikzpicture}
\caption{The \toffca .\label{fig:Toffoli}} Each cell is made of two bits. The leftmost bit at the next time step is just the rightmost bit of the left neighbor at the previous time step ($b$). The rightmost bit at the next time step is the leftmost bit at the previous time step ($c$), inverted if both $b$ and the rightmost bit ($d$) were set to one, i.e. $c\oplus b\cdot d$.
\end{figure}

\begin{figure}[htbp]
\begin{tikzpicture}
\clip (-1.5,-1) rectangle (6.5,4);

\linecircuit{-0.625}{0}
\linecircuit{0.625}{0}
\linecircuit{2.5-0.625}{0}
\linecircuit{2.5+0.625}{0}
\linecircuit{5-0.625}{0}
\transvershaut{1.25-0.625}{0}
\transversbas{1.25+2.5-0.625}{0}

\linecircuit{5+0.625}{0}
\linecircuit{-2.5+0.625}{0}
\transvershaut{1.25+5-0.625}{0}
\transversbas{1.25-2.5-0.625}{0}
\transvershaut{1.25-5-0.625}{0}

\tofcel{0}{0}{$0$}{$0/1$}
\tofcel{2.5}{0}{$-$}{$+$}
\tofcel{5}{0}{$0$}{$0$}

\tofcel{0}{3}{$0$}{$0$}
\tofcel{2.5}{3}{$0/1$}{$-$}
\tofcel{5}{3}{$+/-$}{$0$}
\end{tikzpicture}
\caption{\label{fig:Toffoli_quantique} For the \toffca , the quantum neighborhood $\qneig\tof$ contains $-2$. Indeed, focus on the middle {\sc Toffoli} gate, and forget about its left wire for a second. We are left with a ${\rm {\textsc CNot}}$ gate, apparently controlled by its right wire. But it is a well-known (albeit curious) fact that switching to the $\ket{\pm}$ basis flips who controls whom. So, this ${\rm {\textsc CNot}}$ gate, if activated, will effectively toggle the right wire. Now, recalling that this ${\rm {\textsc CNot}}$ gate is in fact part of a {\sc Toffoli} gate and thus activated by the left wire, we see that changing the left, toggles the right.}
\end{figure}
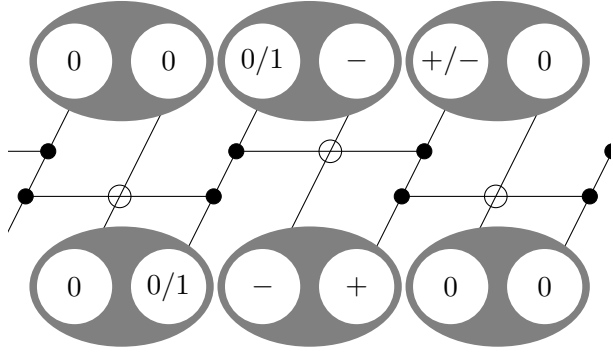

Whenever a transition rule commutes with translations, the neighborhoods have the property that $(x,y)\in\joliN{\Leftrightarrow}(x+z,y+z)\in\joliN$, and are hence completely characterized by the set $\joliN'=\{y-x|(x,y)\in\joliN\}$.  This is the case for the \toffca\, for which we have $\joliN'(\tof)=\acco{-1,0}$. Its inverse neighborhood is also quite small: $\joliN'(\tof^{-1})=\acco{0,1}$. Indeed, each {\sc Toffoli} gate is its own inverse, so $\tof^{-1}$ can be represented simply by turning Figure~\ref{fig:Toffoli} upside-down. So the discrepancy between forward and a possibly much larger inverse neighborhood is irrelevant here. Yet, the quantum neighborhood is strictly larger than its classical neighborhood.
For instance, using the fact that ${\rm {\textsc CNot}}\ket{+-}=\ket{--}$, Figure~\ref{fig:Toffoli_quantique} shows how to transmit information from cell $0$ to cell $2$. This does saturates the upper bounds (\ref{onestep}) and (\ref{onestepDual})
in the sense that the speed in $\qneig{f}$ is least twice that of both $\joliN(f)$ and $\joliN(f^{-1})$. If we look at things more closely, however, the upper bounds (\ref{onestep}) and (\ref{onestepDual}) would both give $\acco{-2,-1,0,1}$, whereas we actually have $\joliN'(Q_f)=\acco{-2,-1,0}$. However it is clear that a symmetrized version of the \toffca\ (two symmetrical versions of it, running in parallel) would saturate the bounds $\{-2\ldots 2\}$ on both sides.

\section{Asymptotics}
\subsection{\toffca : no asymptotic quantum speedup}

\begin{figure}[htbp]
\begin{tikzpicture}[scale=0.6]
\clip (-1.5,-1.3) rectangle (9,7.3);

\linecircuit{-0.625}{0}
\linecircuit{0.625}{0}
\linecircuit{2.5-0.625}{0}
\linecircuit{2.5+0.625}{0}
\transvershaut{1.25-0.625}{0}
\linecircuit{-2.5+0.625}{0}
\transversbas{1.25-2.5-0.625}{0}

\begin{scope}[color=lightgray]
\transvershaut{1.25-5-0.625}{0}
\linecircuit{5-0.625}{0}
\linecircuit{5+0.625}{0}
\linecircuit{7.5-0.625}{0}
\linecircuit{7.5+0.625}{0}
\transversbas{1.25+7.5-0.625}{0}
\transvershaut{1.25+5-0.625}{0}
\transversbas{1.25+2.5-0.625}{0}
\end{scope}

\linecircuit{-2.5+0.625}{3}
\linecircuit{-0.625}{3}
\linecircuit{0.625}{3}
\linecircuit{2.5-0.625}{3}
\linecircuit{2.5+0.625}{3}
\linecircuit{5-0.625}{3}
\transvershaut{1.25-0.625}{3}
\transversbas{1.25+2.5-0.625}{3}
\linecircuit{5+0.625}{3}

\transversbas{1.25-2.5-0.625}{3}

\begin{scope}[color=lightgray]
\linecircuit{7.5-0.625}{3}
\linecircuit{7.5+0.625}{3}
\transvershaut{1.25-5-0.625}{3}
\transvershaut{1.25+5-0.625}{3}
\transversbas{1.25+7.5-0.625}{3}
\end{scope}

\tofcel{0}{0}{}{}
\draw[->,ultra thick] (0,-1.3)--(0,-0.8);

\tofcel{2.5}{0}{}{}
\tofcel{5}{0}{}{}
\tofcel{7.5}{0}{}{}

\tofcel{0}{3}{}{}
\tofcel{2.5}{3}{}{}
\tofcel{5}{3}{}{}
\tofcel{7.5}{3}{}{}

\tofcel{0}{6}{}{}
\tofcel{2.5}{6}{}{}
\tofcel{5}{6}{}{}
\tofcel{7.5}{6}{}{}
\draw[->,ultra thick] (0,6.8)--(0,7.3);
\draw[->,ultra thick] (2.5,6.8)--(2.5,7.3);
\draw[->,ultra thick] (5,6.8)--(5,7.3);
\draw[->,ultra thick] (7.5,6.8)--(7.5,7.3);
\end{tikzpicture}
~~~~~~
\begin{tikzpicture}[scale=0.6]
\clip (-1.5,-1.3) rectangle (9,7.3);

\linecircuit{-0.625}{0}
\linecircuit{0.625}{0}
\linecircuit{2.5-0.625}{0}
\linecircuit{2.5+0.625}{0}
\transvershaut{1.25-0.625}{0}
\linecircuit{-2.5+0.625}{0}
\transversbas{1.25-2.5-0.625}{0}

\begin{scope}[color=lightgray]
\transvershaut{1.25-5-0.625}{0}
\linecircuit{5-0.625}{0}
\linecircuit{5+0.625}{0}
\linecircuit{7.5-0.625}{0}
\linecircuit{7.5+0.625}{0}
\transversbas{1.25+7.5-0.625}{0}
\transvershaut{1.25+5-0.625}{0} %
\transversbas{1.25+2.5-0.625}{0}
\end{scope}

\linecircuit{-2.5+0.625}{3}
\linecircuit{-0.625}{3}
\linecircuit{0.625}{3}
\linecircuit{2.5-0.625}{3}
\linecircuit{2.5+0.625}{3}
\linecircuit{5-0.625}{3}
\transvershaut{1.25-0.625}{3}
\transvershaut{1.25+2.5-0.625}{3} %
\linecircuit{5+0.625}{3}

\transversbas{1.25-2.5-0.625}{3}

\begin{scope}[color=lightgray]
\linecircuit{7.5-0.625}{3}
\linecircuit{7.5+0.625}{3}
\transvershaut{1.25-5-0.625}{3}
\transversbas{1.25+5-0.625}{3}
\transversbas{1.25+7.5-0.625}{3}
\end{scope}

\tofcel{0}{0}{}{}
\draw[->,ultra thick] (0,-1.3)--(0,-0.8);

\tofcel{2.5}{0}{}{}
\tofcel{5}{0}{}{}
\tofcel{7.5}{0}{}{}

\tofcel{0}{3}{}{}
\tofcel{2.5}{3}{}{}
\tofcel{5}{3}{}{}
\tofcel{7.5}{3}{}{}

\tofcel{0}{6}{}{}
\tofcel{2.5}{6}{}{}
\tofcel{5}{6}{}{}
\tofcel{7.5}{6}{}{}
\draw[->,ultra thick] (0,6.8)--(0,7.3);
\draw[->,ultra thick] (2.5,6.8)--(2.5,7.3);
\draw[->,ultra thick] (5,6.8)--(5,7.3);
\draw[->,ultra thick] (7.5,6.8)--(7.5,7.3);
\end{tikzpicture}
\caption{The quantum neighborhood of the iterated \toffca\ does not grow very fast. The bottom-left cell can spread its influence only through the black gates ({\em left}). In the first step, because it touches the left wire of a {\sc Toffoli} gate, it can signal at speed two. But it only reaches the rightmost bit of the third cell, so that in the second step, because the top-left {\sc Toffoli} gates commute ({\em right}), there is no way it can signal at speed two again.\label{fig:Toffoli_asymptotics}}
\end{figure}
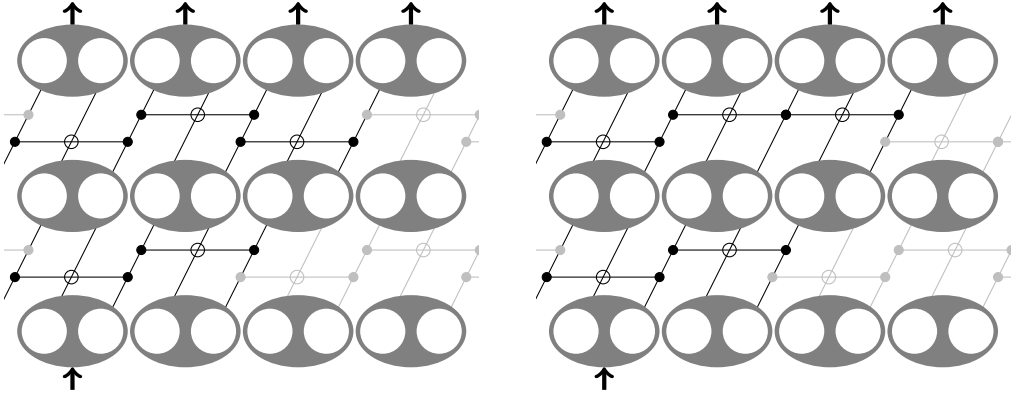

It would thus appear that quantum information can travel twice as fast as classical information under the \toffca. But in order to realize this gain, we must iterate the automaton.  Figure \ref{fig:Toffoli_asymptotics} shows the result. The iterated quantum neighborhood is again larger than its classical counterpart, but not by much: it is $\qneig{f^n}=\acco{-(n+1),-n,\ldots,0}$, which differs from $\joliN (f^n)$ by the single cell $-(n+1)$.
Indeed, we have seen how in one step the right component of cell $-2$ reaches the left component of cell $0$. But then, as the \textsc{Toffoli} gates commute, the left component of cell $0$ can only reach the left component of cell $1$, and so on.

The \toffca\ supports a large first step, but from then on no further speedup---is this a special, or a general property? The following corollary settles this question. It comes from and Proposition \ref{prop:main} and proposition 2.3 in \cite{block}, its proof is to be found in \cite{block}, and is too tedious to be reproduced here.

\begin{cor} \label{prop:iter}
$\qneig{f_n\cdots f_1} \incl \bigcup\limits_{k=1}^n \joliNT\pa{(f_n \cdots f_{k+1})^{-1}}\qneig{f_k}\joliN(f_{k-1}\cdots f_{1})$.
\end{cor}

If we apply this result to the case where $f_1=f_2=\ldots=f_n=f$, we get 

\begin{equation}
\qneig{f^n} \incl \bigcup\limits_{k=1}^n \joliNT (f^{k-n})\qneig{f}\joliN(f^{k-1})
\end{equation}

When $f$ is the \toffca ,  we have $\joliN(f^n)=\joliNT(f^{-n})$, which implies that all the quantum speedup of $\qneig{f^n}$ just comes from the single-step speedup: The ratio between the sizes of $\qneig{f^n}$ and $\joliN(f^n)$ asymptotically goes to $1$, as $n$ goes to infinity.

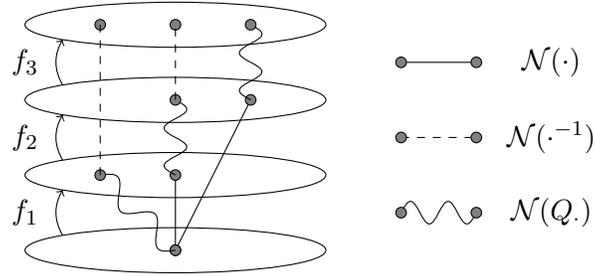
\begin{figure}[htbp]
\begin{tikzpicture}
\tikzstyle{gloubi}=[circle, draw, fill=black!50,
                        inner sep=0pt, minimum width=4pt]

\draw (0,0) ellipse  (2cm and 0.3cm);
\draw (0,1) ellipse  (2cm and 0.3cm);
\draw (0,2) ellipse  (2cm and 0.3cm);
\draw (0,3) ellipse  (2cm and 0.3cm);

\draw (0,0) node[gloubi] {} -- (0,1) node[gloubi] {};
\draw (0,0) node[gloubi] {} -- (1,2) node[gloubi] {};

\draw[decoration={
            complete sines,
            segment length=0.5cm,
            amplitude=0.3cm
        },
        decorate] 
 (0,0) -- (-1,1) node[gloubi] {};

\draw[decoration={
            complete sines,
            segment length=0.5cm,
            amplitude=0.3cm
        },
        decorate] 
 (0,1) -- (0,2) node[gloubi] {};
\draw[decoration={
            complete sines,
            segment length=0.5cm,
            amplitude=0.3cm
        },
        decorate] 
 (1,2) -- (1,3) node[gloubi] {};

\draw (-1,3) node [gloubi] {} ;
\draw (0,3) node [gloubi] {} ;

\draw[dashed] (-1,1) {} -- (-1,3) {} ;
\draw[dashed] (0,2) {} -- (0,3)  {} ;

\draw (3,2.5) node[gloubi] {} -- (4,2.5) node[gloubi] {} ;
\draw (5,2.5) node {$\joliN(\cdot)$} ;

\draw[dashed] (3,1.5) -- (4,1.5) ;
\draw (3,1.5) node[gloubi] {};
\draw (4,1.5) node[gloubi] {};
\draw (5,1.5) node {$\joliN(\cdot^{-1})$} ;

\draw%
[decoration={
            complete sines,
            segment length=0.5cm,
            amplitude=0.3cm
        },
        decorate]
(3,0.5) -- (4,0.5) ;

\draw (3,0.5) node[gloubi] {};
\draw (4,0.5) node[gloubi] {};
\draw (5,0.5) node {$\qneig {\cdot}$} ;

\draw[->=latex] (-1.5,0.2) to [bend left=30] (-1.5,0.8)  ;
\draw (-2,0.5) node {$f_1$} ;
\draw[->=latex] (-1.5,1.2) to [bend left=30] (-1.5,1.8) ;
\draw (-2,1.5) node {$f_2$} ;
\draw[->=latex] (-1.5,2.2) to [bend left=30] (-1.5,2.8) ;
\draw (-2,2.5) node {$f_3$} ;

\end{tikzpicture}
\caption{\label{fig:iter} Illustration of Corollary~\ref{prop:iter} when $n=3$.}
\end{figure}

In the general case the result involves a composition of arbitrary $f_i$-s, and is illustrated in Figure~\ref{fig:iter} for arbitrary $f_i$-s and $n=3$. 
Although it is difficult to give an intuitive explanation as to \emph{why} this corollary is true, it is not too hard to grasp what it \emph{means}: The  messages that can be transmitted from one location to another in a quantum universe ruled by the dynamics $f_n\cdots f_1$ must follow a particular protocol.  First, they are transmitted trough $k-1$ steps under a purely classical form --- this is denoted in Figure~\ref{fig:iter} with a plain line.  Then, on the $k$-th step, similarly to the trick we used for the \toffca, something quantum happens --- this is denoted with a wavy line.  Afterwards, through the remaining $n-k$ steps, this message is transmitted in a way that can be described as dual to a classical channel --- this is denoted with a dotted line.

\begin{figure}[htbp]
  \label{fig_diagramme}
  \begin{tikzpicture}[
        scale=1.5,
        axis/.style={very thick, ->, >=stealth'},
        pointille/.style={dashed},
     ]
    \coordinate (origine) at (0,0) ;
    \coordinate (hautdroite) at (1,2.5) ;
    \coordinate (hautgauche) at (-1.4,2.5) ;
    \coordinate (basgauche) at (-0.3,-2.5) ;
	\coordinate (basdroite) at (1.8,-2.5) ;
	\coordinate (basgg) at (-1.5,-2.5) ;
	\coordinate (basdd) at (2.3,-2.5) ;
	\coordinate (hautgg) at (-2.3,2.5) ;
	\coordinate (hautdd) at (1.5,2.5) ;

    \def\secteur{(-3,1) -- (-3,-1) -- (origine) -- (3,1) -- (3,-1) -- (origine) } ;
    \def\tout{(-2.5,-2.5) rectangle (2.5,2.5)} ;

    \draw[axis] (-2.5,0) -- (2.5,0) node(xline) [right] {space} ;
    \draw[axis] (0,-3) -- (0,3) node(xline) [above] {time} ;
	\fill[gray, path fading=north] (origine) -- (hautdroite) -- (hautgauche) ;
	\fill[gray, path fading=south] (origine) -- (basdroite) -- (basgauche) ;
\begin{scope}[even odd rule]
		\clip  \secteur \tout ;
	\fill[lightgray, path fading=north] (origine) -- (0.5,0) -- (hautdd) -- (hautdroite) ;
\end{scope}
\begin{scope}[even odd rule]
		\clip  \secteur \tout ;
	\fill[lightgray, path fading=south] (origine) -- (0.5,0) -- (basdd) -- (basdroite) ;
\end{scope}	
\begin{scope}[even odd rule]
		\clip  \secteur \tout ;
	\fill[lightgray, path fading=south] (origine) -- (-0.5,0) -- (basgg) -- (basgauche) ;
\end{scope}
	\begin{scope}[even odd rule]
		\clip  \secteur \tout ;
	        \fill[lightgray, path fading=north] (origine) -- (-0.5,0) -- (hautgg) -- (hautgauche)  ;
	\end{scope}
	\draw[pointille] (origine) -- (-1.8,2.5) ;
	\draw[pointille] (origine) -- (-1,-2.5) ;
	\node (pncan) at (-0.13,1.5) {classical future};
	\node (pncan) at (0.4,-1.5) {classical past};
	\node (pncan) at (-0.1,0.5) {quantum future};
	\node (pncan) at (0.1,-0.5) {quantum past};
   \end{tikzpicture}
  \caption{Typical asymptotics for iterated dynamics.}
  \label{fig:spot}
\end{figure}
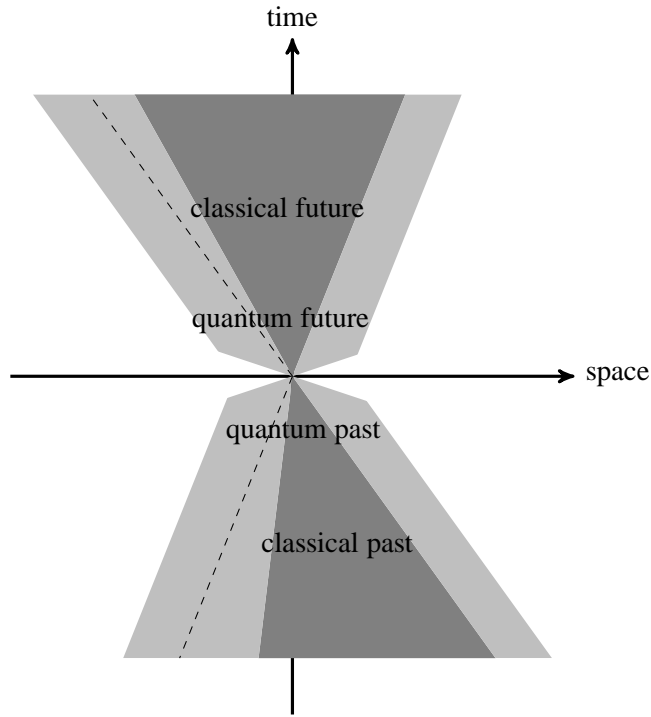

Figure \ref{fig:spot}, on the other hand, illustrates it when all of the $f_i$-s are equal, for an arbitrary number of steps of a one-dimensional CA. Notice that the dotted lines in this Figure \ref{fig:spot} suggest that, in the specific case when $\joliN_{f^{-1}}\supset\joliN(f)$, the corresponding QCA $Q_f$ could, in principle, take that one benefit to signal asymptotically faster than $f$. Since this is not the case of the \toffca\ we now need an example showing that when $\joliN(f^{-n})$ is larger than $\joliN(f^n)$, you can actually transmit information at a long distance, thereby saturating Corollary \ref{prop:iter}.

\subsection{\textsc{J-CA}: achieving the maximal quantum speed}

Let $\Sigma=\pa{\Z/2\Z}^d$. For $x\in\Sigma$, $x^i$ denotes its $i$-th component. We define a CA $J_d$ on this cell structure in the following way:

\begin{equation}
J_d(v)_0^i=\left\{ \begin{array}{ll}
v_0^i+v_1^{i+1}& \text{if $i<d$} \\
v_1^1 &  \text{if $i=d$}
\end{array}\right..
\end{equation}

Its inverse is given by 
\begin{equation}
J_d^{-1}(v)_0^i = v_{-i}^d + \sum\limits_{j=1}^{i-1} v_{j-i}^j.
\end{equation}

We thus have $\joliN(J_d)=\acco{0,1}$ and $\joliN(J_d^{-1})=\acco{-d,\ldots,-1}$.  Let us now illustrate the left inclusion of (\ref{onestepDual}) by showing that the quantum neighborhood of $J_d$ does indeed contain $d$.

Let $b$ be the zero configuration defined by $b_k=(0,\ldots,0)$ for all $n\in\Z$, and $c$ defined by $c_k^i=\delta_{in}$. It is easily checked that $f(b)$ is also the zero word, whereas $f(c)_k^i=\delta_{k0}$. The point is that $f(b)$ and $f(c)$ coincide on $\Z\setminus\acco{0}$, whereas $b_d\neq c_d$. Now, imagine Alice and Bob respectively live in cell $d$ and $0$, and that the dynamics of their universe is described by $Q(J_d)$.  Assuming they share prior entanglement, Alice can transmit a message to Bob in a single step.  Indeed, say the initial state is $\ket{\psi_+}=\frac{1}{\sqrt{2}}\pa{\ket{b}+\ket{c}}$. Since Alice is at a place where she can distinguish $b$ from $c$, she can, by applying a local controlled phase gate, switch at will from $\ket{\psi_+}$ to $\ket{\psi_-}=\frac{1}{\sqrt{2}}\pa{\ket{b}-\ket{c}}$. One time step later, their world is in the pure state $\ket{\phi_{\pm}}=\frac{1}{\sqrt{2}}\pa{\ket{J_d(b)}\pm \ket{J_d(c)}}$. Since $J_d(b)$ and $J_d(c)$ coincide outside Bob's place, where they are equal to zero, one can write $\ket{\phi_{\pm}}=\ket{\underline{0}}\otimes \ket{\varphi_{\pm}}$, where $\ket{\varphi_\pm}$ is totally accessible to Bob, and so that he can easily observe, with local measurement, whether Alice switched from $\ket{\psi_+}$ to $\ket{\psi_-}$. Alice was thus able to transmit one bit of information to Bob in just one time step, proving $d\in \qneig{J_d}$.

As for the asymptotic bound, $\joliN(J_d^n)$ must be included in $\joliN(J_d)^n=\acco{0,\ldots,n}$, but one can notice, for instance, that $J_d^{-n}(v)_0^d$ contains the term $v_{-dn}^d$ also exactly once; Therefore $dn\in\joliNT(J_d^{-n})$.  Hence this example saturates the asymptotic upper bound. It is a pure example where the sound cone of the quantized automaton is just the union of the classical sound cone ($\joliN(f^n)$) and its dual ($\joliNT(f^{-n})$).  As $n$ increases, the ratio of the widths of the classical cone and the quantum one remains a constant $d$.  This example can also be easily symmetrized, by taking $\Sigma=\pa{\Z/2\Z}^{2d}$, applying $J_d$ on the first $d$ entries, and the symmetrized of $J_d$ on the others.

\section*{Conclusion}

Let us summarize the main points of interest.
\begin{itemize}

\item For a single step of a dynamics $f$, quantum information can jump unboundedly further than classical information, as propagated by $f$. This typically requires prior entanglement shared between parties.

\item Quantum information can also jump further than classical information, as propagated by both $f$ and $f^{-1}$, but only in a bounded way. We give optimal bounds on this quantum neighborhood, as a function of the neighborhoods of $f$ and $f^{-1}$.

\item Therefore, even though the neighborhood of $f^{-1}$ cannot be bounded
by a computable function in terms of the neighborhood of $f$ \cite{KariRevUndec,undecidable,Kari_blocks}, it is still the case that if we are given both the neighborhoods of $f$ and $f^{-1}$, then we can bound the quantum neighborhood.

\item When iterating the dynamics $f$, quantum information can again flow asymptotically unboundedly faster than classical information, as propagated by $f$. But it cannot flow asymptotically faster than classical information as propagated by both $f$ and $f^{-1}$.

\item Therefore in the case of an evolution with a proper time symmetry, quantum information cannot flow asymptotically faster than classical information.
\end{itemize}

Future works include of course a better understanding and physical interpretation of this channel, which should make full use of the duality.

\section*{Acknowledgements}
This work has been funded by the ANR-12-BS02-007-01 TARMAC grant, the ANR-10-JCJC-0208 CausaQ grant, and the Deutsche Forschungsgemeinschaft (Forschergruppe 635). This work has been partially done at IXXI. 

\bibliographystyle{acm}


\appendix*

\section{The quantum neighborhood contains the block neighborhood}
\begin{prop}
$\BN(f)\incl \joliN(Q_f)$.
\end{prop}

Let $x,y\in X$ such that $(x,y)\not\in \joliN(Q_f)$, i.e. such that $Q_f(\joliA_{y})\incl \joliA_{X\setminus \acco{x}}$.  We have to prove that $(x,y)\not\in\BN(f)$, i.e. that $f$ can be semilocalized as in Figure~\ref{fig:semilocal} with $Z=X\setminus\acco{x}$.  We will proceed in two steps. In the first step, we prove four combinatorial properties that we will tap into in the second and final step, where we explicitly construct the bijections $g$ and $h$ as seen in Figure~\ref{fig:semilocal}.

\noindent{\bf \em Four properties}

We will prove that these hold:

\begin{itemize}
\item[(1)] $(x,y)\not\in\joliN(f)$;
\item[(2)] $(y,x)\not\in\joliN(f^{-1})$;
\item[(3)] When $v,w\in\joliC$ are such that $v_{x}=w_{x}$, it is enough to know $v_{X\setminus\acco{x}}$ and $w_{X\setminus\acco{x}}$ in order to determine whether $f(v)_{X\setminus \acco{y}}=f(w)_{X\setminus \acco{y}}$;
\item[(4)] When $v,w\in\joliC$ are such that $f(v)_{y}=f(w)_{y}$, it is enough to know $f(v)_{X\setminus\acco{y}}$ and $f(w)_{X\setminus\acco{y}}$ in order to determine whether $v_{X\setminus \acco{x}}=w_{X\setminus \acco{x}}$.
\end{itemize}

In order to do so, let us introduce some notations.  For $v,w\in\joliC$ and $a,b\in \Sigma_{y}$, let $q(v,w,a,b)$ be $\bra{v}Q_f\pa{\ket{a}\bra{b}} \ket{w}$.  Since $q(v,w,a,b)=\sum\limits_{u\in\Sigma_{X\setminus\acco{y}}}\bra{f(v)} (\ket{a}\bra{b} \otimes \ket{u}\bra{u})\ket{f(w)}$, we have 

\begin{equation}
q(v,w,a,b)=\left\{ 
\begin{array}{ll}
1 & \text{if $f(v)_{y}=a$, $f(w)_{y}=b$ and $f(v)_{X\setminus \acco{y}}=f(w)_{X\setminus \acco{y}}$} \\ 
0 & \text{otherwise}
\end{array}
\right.
\end{equation}

Since $Q_f\pa{\joliA_{y}}\incl\joliA_{X\setminus\acco{x}}$, we have $Q_f\pa{\joliA_{y}}=M\otimes \mathbb{I}_{\joliA_{x}}$ for some $M\in\joliA_{X\setminus\acco{x}}$, from which we deduce that $q$ has the following properties :

\begin{itemize}
\item[(i)] if $v_{x}\neq w_{x}$, then $q(v,w,a,b)=0$;
\item[(ii)] if $v_{x}= w_{x}$, then $q(v,w,a,b)$ depends only on $v_{X\setminus\acco{x}}$ and $w_{X\setminus\acco{x}}$, $a$ and $b$.
\end{itemize}

Let us now prove points (1), (2), (3) and (4).

\begin{itemize}
\item[(1)] In order to prove that $x$ is not in the classical neighborhood of $y$ for $f$, we have to show that, for any configurations $v$ and $w$ that coincide on $X\setminus\acco{x}$, $f(v)_{y}=f(w)_{y}$.  Let then $v,w\in\joliC$ such that $v_{X\setminus\acco{x}}=w_{X\setminus\acco{x}}$.  First, $q\pa{v,v,f(v)_{y},f(v)_{y})}=1$.  But, by (ii), since $w_{X\setminus\acco{x}}=v_{X\setminus\acco{x}}$, we get $q\pa{w,w,f(v)_{y},f(v)_{y}}=1$, which means $f(v)_{y}=f(w)_{y}$.

\item[(2)] Similarly, in order to prove that $y$ is not in the classical neighborhood of $x$ for $f^{-1}$, we have to show that, for any configurations $v$ and $w$ such that $f(v)$ and $f(w)$ coincide on $X\setminus\acco{y}$, $v_{x}=w_{x}$.  Let then $v,w\in\joliC$ such that $f(v)_{X\setminus\acco{y}}=f(w)_{X\setminus\acco{y}}$.  We then have $q\pa{v,w,f(v)_{y},f(w)_{y}}=1$, which implies, by (i), that $v_x=w_x$.

\item[(3)] Let $v,w \in \joliC$ be configurations such that $v_{x}=w_{x}$. Then $q(v,w,f(v)_y,f(w)_y)=1$ if and only if $f(v)_{X\setminus \acco{y}}=f(w)_{X\setminus \acco{y}}$. Can this quantity be determined knowing only $v_{X\setminus\acco{x}}$ and $w_{X\setminus\acco{x}}$? Yes: We have already proven in (1) that $f(v)_y$ and $f(w)_y$ are determined by $v_{X\setminus\acco{x}}$ and $w_{X\setminus\acco{x}}$; and by hypothesis, for any fixed $a,b\in\Sigma_y$, $q(v,w,a,b)$ depends only on $v_{X\setminus\acco{x}}$ and $w_{X\setminus\acco{x}}$.

\item[(4)]  We want to prove that, for any $v,w \in \joliC$ such that $f(v)_{y}=f(w)_{y}$, it is enough to know $f(v)_{X\setminus\acco{y}}$ and $f(w)_{X\setminus\acco{y}}$ to determine whether $v_{X\setminus \acco{x}}=w_{X\setminus \acco{x}}$.

In order to do so, let $v,w,v',w'\in\joliC$ be such that $f(v)_{y}=f(w)_{y}$, $f(v')_{y}=f(w')_{y}$, $f(v)_{X\setminus\acco{y}}=f(v')_{X\setminus\acco{y}}$, $f(w)_{X\setminus\acco{y}}=f(w')_{X\setminus\acco{y}}$ and $v_{X\setminus \acco{x}}=w_{X\setminus \acco{x}}$. We have to prove that  $v'_{X\setminus \acco{x}}=w'_{X\setminus \acco{x}}$.

First, from (2) and $f(v)_{X\setminus\acco{y}}=f(v')_{X\setminus\acco{y}}$, we get $v_x=v'_x$ ; therefore, according to (3), not only do we have $f(v_{X\setminus\acco{x}}.v_x)_{X\setminus\acco{y}}=f(v'_{X\setminus\acco{x}}.v_x)_{X\setminus\acco{y}}$, but for any $a\in\Sigma_x$, $f(v_{X\setminus\acco{x}}.a)_{X\setminus\acco{y}}=f(v'_{X\setminus\acco{x}}.a)_{X\setminus\acco{y}}$.  Likewise, for any $a\in\Sigma_x$, $f(w_{X\setminus\acco{x}}.a)_{X\setminus\acco{y}}=f(w'_{X\setminus\acco{x}}.a)_{X\setminus\acco{y}}$.

Let $a$ be an arbitrary element of $\Sigma_x$.   
Since, by assumption, $v_{X\setminus \acco{x}}=w_{X\setminus \acco{x}}$, we can therefore deduce the following :

$$f(v'_{X\setminus\acco{x}}.a)_{X\setminus\acco{y}}=f(v_{X\setminus\acco{x}}.a)_{X\setminus\acco{y}}=f(w_{X\setminus\acco{x}}.a)_{X\setminus\acco{y}}=f(w'_{X\setminus\acco{x}}.a)_{X\setminus\acco{y}}.$$

Moreover, from (1) we get $f(v'_{X\setminus\acco{x}}.a)_{y}=f(w'_{X\setminus\acco{x}}.a)_{y}$.  Therefore $f(v'_{X\setminus\acco{x}}.a)=f(w'_{X\setminus\acco{x}}.a)$; since $f$ is one-to-one, we conclude that $v'_{X\setminus\acco{x}}=w'_{X\setminus\acco{x}}$.
\end{itemize}

\noindent {\bf \em Block construction}

Let $\sim_x$ be the binary relation on $\Sigma_{X\setminus\acco{x}}$ defined by 

$$v\sim_x v' \quad \text{iff} \quad \forall a\in \Sigma_x\;f(v.a)_{X\setminus\acco{y}}=f(v'.a)_{X\setminus\acco{y}}.$$

Note that, because of (3), this is actually equivalent to $\exists a\in \Sigma_x\;f(v.a)_{X\setminus\acco{y}}=f(v'.a)_{X\setminus\acco{y}}$, from which we deduce
\begin{equation}
\forall b,b'\in \Sigma_y\;\forall w\in \Sigma_{X\setminus\acco{y}}\quad  f^{-1}(w.b)_{X\setminus\acco{x}}\sim_x f^{-1}(w.b')_{X\setminus\acco{x}} \label{existent}
\end{equation}

Indeed, given any $b,b'\in \Sigma_y$ and $w\in \Sigma_{X\setminus\acco{y}}$, we can set $v=f^{-1}(w.b)_{X\setminus\acco{x}}$, $v'=f^{-1}(w.b')_{X\setminus\acco{x}}$ and $a=f^{-1}(w.b)_x$.  Because of (2), we also have $a=f^{-1}(w.b')$, so that $f(v.a)_{X\setminus\acco{y}}=f(v'.a)_{X\setminus\acco{y}}$.

$\sim_x$ is clearly an equivalence relation, so using (1) we can define 

$$\lambda:\left(\begin{array}{rcl} \Sigma_{X\setminus\acco{x}}& \to & \Sigma_y\times (\Sigma_{X\setminus\acco{x}}/\sim_x)\\ v&\mapsto&(f(v.a)_y, [v])\\ \end{array} \right)$$

where $a$ is an arbitrary element of $\Sigma_{x}$ and $[v]$ is the class of $v$ in $\Sigma_{X\setminus\acco{x}}/\sim_x$.   Thanks to (2) and (4), one can likewise define $\sim_y$ on $\Sigma_{X\setminus\acco{y}}$; we have the corresponding property
\begin{equation}
\forall a,a'\in \Sigma_x\;\forall v\in \Sigma_{X\setminus\acco{x}}\quad  f(v.a)_{X\setminus\acco{y}}\sim_y f(v.a')_{X\setminus\acco{y}} \label{existent_dual}
\end{equation}

and we can define

$$\mu:\left(\begin{array}{rcl} \Sigma_{X\setminus\acco{y}}& \to & \Sigma_x\times (\Sigma_{X\setminus\acco{y}}/\sim_y)\\ w&\mapsto &(f^{-1}(w.b)_x, [w])\\ \end{array} \right)$$

where $b$ is an arbitrary element of $\Sigma_y$.

We now introduce $$\alpha:\pa{\begin{array}{rcl} \Sigma_{X\setminus\acco{x}}/\sim_x & \to & \Sigma_{X\setminus\acco{y}}/\sim_y \\ \left[v\right] & \mapsto & \left[f(v.a)_{X\setminus\acco{y}}\right] \end{array}}$$ 

where $a$ is again an arbitrary element of $\Sigma_x$.  Let us prove that $\alpha$ is a well-defined bijection.  In order to prove that it is well-defined, we need to show that for every $v,v'\in X\setminus\acco{x}$ such that $v\sim_x v'$ and every $a,a'\in \Sigma_x$, $f(v.a)_{X\setminus\acco{y}}\sim_y f(v'.a')_{X\setminus\acco{y}}$, which is easily done in two small steps.  First, by definition of $\sim_x$, $f(v.a)_{X\setminus\acco{y}}=f(v'.a)_{X\setminus\acco{y}}$.  Then, by (\ref{existent_dual}), $f(v'.a)_{X\setminus\acco{y}}\sim_y f(v'.a')$.  We now prove that $\alpha$ is bijection by constructing its inverse $\beta$.  We define $\beta$ as follows:

$$\beta:\pa{\begin{array}{rcl} \Sigma_{X\setminus\acco{y}}/\sim_y & \to & \Sigma_{X\setminus\acco{x}}/\sim_x \\ \left[w\right] & \mapsto & \left[f^{-1}(w.b)_{X\setminus\acco{x}}\right] \end{array}}$$

We then have $\beta\alpha ([v])=[f^{-1}(f(v.a)_{X\setminus\acco{y}}.b)_{X\setminus\acco{x}}]$.  Since this value is independent of $b$, we can try in particular with $b=f(v.a)_y$, where it is clear that we get $[v]$.

We near the end of our construction, which consists in finding a block decomposition according to Figure~\ref{fig:semilocal}, where $Z=X\setminus\acco{x}$.  We define $g=(\id_{\Sigma_{y}}\times \alpha) \lambda : \Sigma_{X\setminus\acco{x}}\to \Sigma_y\times (\Sigma_{X\setminus\acco{y}}/\sim_y)$.  It is a bijection because we can define its inverse $g^{-1}$ by $g^{-1}(b,[w])=f^{-1}(w.b)_{X\setminus\acco{x}}$, which is well-defined by definition of $\sim_y$.

Since $\alpha$ and $g$ are bijections, so is $\lambda$, and so is $\mu$ symmetrically.  We can therefore define $h=\mu^{-1}$ and thus complete our proof.

\end{document}